\documentclass[10pt,twocolumn]{revtex4-1}

\usepackage{amsmath, amsfonts, amssymb}     % maths symbols
\usepackage{graphicx}                       % coloured text, subfigs
\usepackage{xfrac}                          % inline fractions
\usepackage{grffile}                        % filenames with dots
\usepackage{color,colortbl}

\usepackage[colorlinks, citecolor=blue, urlcolor=red, linkcolor=blue]{hyperref}

\renewcommand{\eqref}{Eq.~\ref}

\newcommand{\gdot}[0] {\dot{\gamma}}
\newcommand{\sigmay}[0] {\sigma_{\rm y}}
\newcommand{\gdotbar}[0] {\bar{\dot{\gamma}}}

\newcommand{\xhat}{\hat{\mathbf{x}}}
\newcommand{\yhat}{\hat{\mathbf{y}}}

\newcommand{\tens}[1]{\mathbf{{#1}}}

\newcommand{\tw}{t_{\rm w}}

\newcommand{\be}{\begin{equation}}
\newcommand{\ee}{\end{equation}}
\newcommand{\bea}{\begin{eqnarray}}
\newcommand{\eea}{\end{eqnarray}}

\begin{document}
\title{Long-term memory and delayed shear localisation in soft glassy materials}
\author{Henry A. Lockwood, Matthew P. Carrington and Suzanne M. Fielding}
\affiliation{Department of Physics, Durham University, Science Laboratories,
  South Road, Durham DH1 3LE, UK}
\date{\today}
\begin{abstract}

We study theoretically the dynamics of soft glassy materials during
the process of stress relaxation following the rapid imposition of a
shear strain. By detailed numerical simulations of a mesoscopic soft
glassy rheology model and three different simplified continuum
fluidity models, we show that a dramatic shear localisation
instability arises, in which the strain field suddenly becomes
heterogeneous within the sample, accompanied by a precipitous drop in
the stress. Remarkably, this instability can arise at extremely long
delay times after the strain was applied, due to the long-term memory
inherent to glassy systems. The finding that a catastrophic mechanical
instability can arise long after any deformation could have far
reaching consequences for material processing and performance, and
potentially also for delayed geophysical phenomena.

\end{abstract}
\date{\today}
\maketitle

Many soft materials, including dense colloids, microgels, emulsions
and foams, show notable shared features in their rheological
(deformation and flow) properties.  Their steady state flow curve of
shear stress $\sigma(\gdot)$ as a function of shear rate $\gdot$ shows
a yield stress in the limit of slow flows, $\sigmay=\lim_{\gdot\to
0}\sigma(\gdot)$~\cite{bonn2017yield}. The viscoelastic spectra
$G^*(\omega)$ characterising their stress response to an imposed
strain oscillation are typically rather flat functions of the
oscillation frequency $\omega$~\cite{vlassopoulos2014tunable}. These
shared rheological features indicate the presence of sluggish stress
relaxation modes, and have been attributed to the underlying presence
of the basic glassy features of disorder (e.g., in a disordered packing
of emulsion droplets), and metastability (with large energy barriers
impeding droplet rearrangements)~\cite{sollich1997rheology}.
Similarly amorphous, but harder materials include polymeric, metallic
and structural glasses.

Following the switch-on of a shear flow in an initially well rested
sample, such a material will typically respond initially elastically,
before plastically yielding into a finally fluidised
state~\cite{bonn2017yield}. Commonly observed during this process of
yielding is the phenomenon of {\em shear localisation}: a state of
initially homogeneous shear in the elastic regime gives way during
plastic yielding to the formation of shear
bands~\cite{divoux2010transient,martin2012transient,gibaud2008influence,dimitriou2014comprehensive,colombo2014stress,shi2007evaluation,shrivastav2016heterogeneous,fielding2014shear,moorcroft2011age,manning2007strain,manning2009rate,hinkle2016small,jagla2010shear}:
layers of differing viscosity that coexist within the material. These
bands may eventually heal to leave a homogeneously fluidised flowing
state. In harder materials, shear localisation more often results in
catastrophic material failure~\cite{doyle1972fracture}. In geophysics,
it is implicated in earthquakes, landslides and mudslips~\cite{daub2010pulse,coussot2002avalanche}.

Besides sustained deformations of the kind just described, in which
(given a constant imposed shear rate $\gdot$) the shear strain
$\gamma=\gdot t$ accumulates indefinitely over time $t$ and the
material yields into a steadily flowing state, another commonly
imposed type of deformation involves instead simply straining a
material by a finite amount, which we shall denote $\gamma_0$ in what
follows. The strain is held constant thereafter, with no further
deformation applied. This will be modelled below by a step function,
$\gamma(t)=\gamma_0\Theta(t-\tw)$, though the physics we present holds
for any reasonably short time interval of deformation. We denote the
time of strain application $t=\tw$, defined relative to the sample
having been freshly prepared at an earlier time $t=0$. The shear
stress initially generated by this deformation, $\sigma(t=\tw^+)$,
then typically decays as a function of the subsequent time interval
$\Delta t=t-\tw$, with the material slowly relaxing towards a
stress-free, quiescent state as $\Delta t\to\infty$. Widely observed
in soft glassy materials is the phenomenon of {\em ageing}, in which a
significant part of this stress relaxation takes place on timescales
that grow with the sample age
$\tw$~\cite{suman2018microstructure,kaushal2014linear,rogers2010time,ramos2001ultraslow,yin2008soft,derec2003aging,derec2000rheological,fielding2000aging}:
a property that has been termed `long-term
memory'~\cite{bouchaud1998out}.

Given the absence in such a scenario of any finally flowing state, it
has been widely assumed that this post-strain stress relaxation will
take place in a straightforwardly innocuous way, with the material
simply slowly returning to a homogeneous relaxed state. This Letter
will show, on the contrary, that for a wide range of values of
amplitude $\gamma_0$ and time $\tw$ of imposed strain, the material
will instead suffer a catastrophic internal instability in which it
suddenly becomes highly heterogeneous within itself, accompanied by a
precipitous drop in the stress. We further show that this instability
can be delayed long into the process of stress relaxation, with the
delay time $\Delta t^*$ increasing linearly with the initial sample
age $\tw$. Remarkably, therefore, the delay time can become
arbitrarily large for old systems, $\tw\to\infty$. An observer lacking
any knowledge of the long historic deformation could thus be caught
entirely unawares by the instability.

We shall demonstrate this phenomenon by detailed numerical simulations
of a mesoscopic soft glassy rheology
model~\cite{sollich1997rheology}. We show it also to hold in a three
different variants of a highly simplified continuum fluidity
model~\cite{moorcroft2011age}.  In thus confirming it to be
independent of the particular constitutive model used, we suggest it
may be generic across amorphous materials, with far reaching
consequences for material processing and performance, and potentially
also for delayed geophysical phenomena such as mudslips and seismic
aftershocks.

\begin{figure}[!t]
  \includegraphics[width=8.5cm]{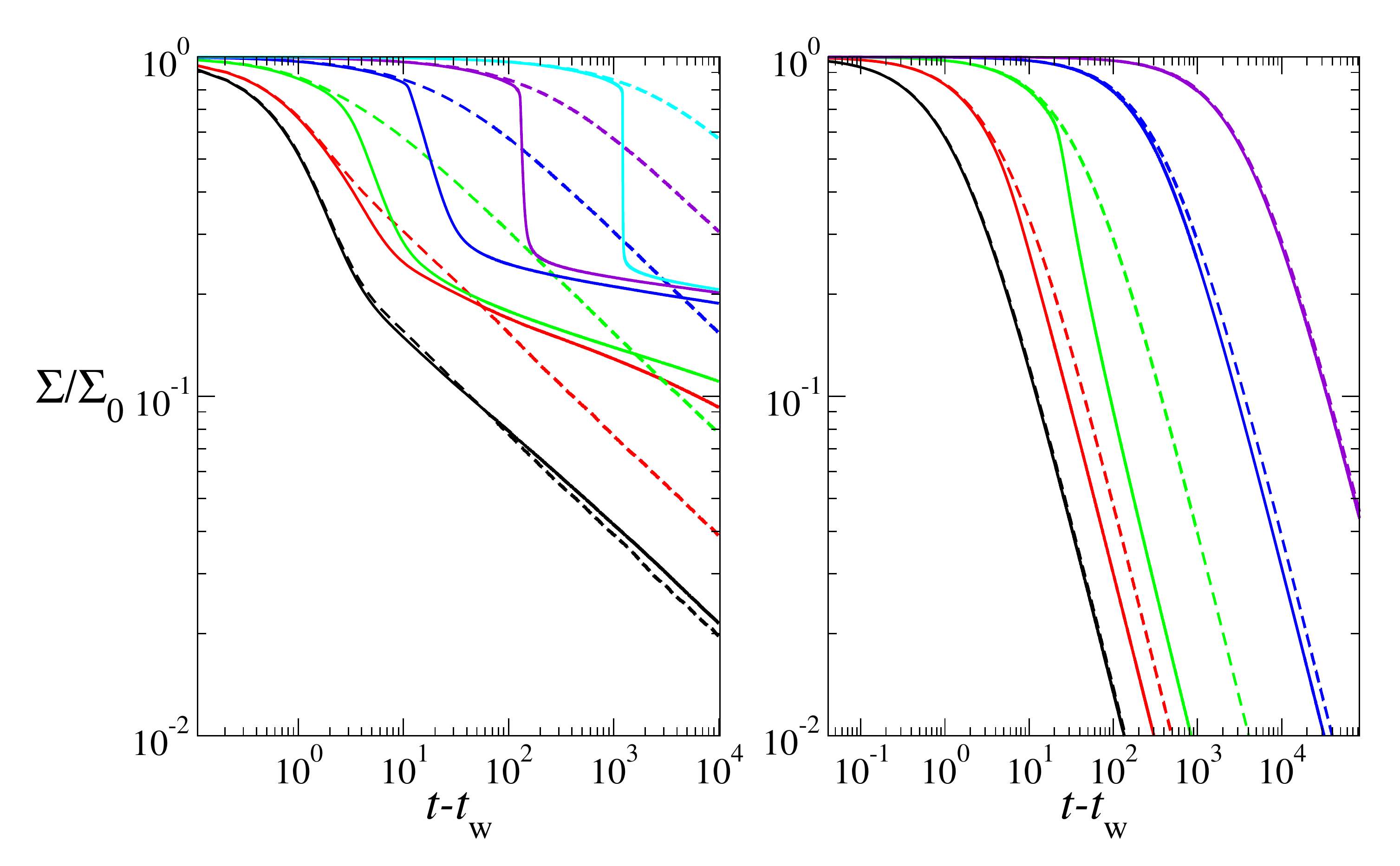} \caption{Stress
  decay as a function of the time interval $\Delta t=t-\tw$ since the
  imposition of a step shear strain. {\bf Left:} SGR model at a noise
  temperature $x=0.3$ for a strain amplitude $\gamma_0=2.5$ and
  waiting times $\tw=10^3,10^4,\cdots 10^8$ in curves left to
  right. {\bf Right:} fluidity model for $\gamma_0=5.5$ and waiting
  times $\tw=10^4,10^5,10^6,10^7,10^8$ in curves left to right. Dashed curves show the results of calculations in which
  the strain field is artificially constrained to remain homogeneous,
  and solid lines in which it is allowed to become heterogeneous.
  \label{fig:fig1}}
\end{figure}

Throughout we assume incompresssible, inertialess deformations in which
the displacement, velocity and stress fields within the material,
$\textbf{u}(\textbf{r},t)$, $\textbf{v}(\textbf{r},t)$ and
$\mathbf{\Sigma}(\textbf{r},t)$, obey the standard conditions of mass
balance, $\nabla.\textbf{u}=0$ and $\nabla.\textbf{v}=0$, and of force
balance, $\nabla.\mathbf{\Sigma} = 0$. The total stress
$\mathbf{\Sigma}=
\boldsymbol{\sigma} + 2\eta\tens{D} -p\tens{I}$ in any fluid element
is assumed to comprise an elastoplastic contribution $\boldsymbol{\sigma}$
from the mesoscopic substructures (emulsions droplets, microgel beads,
etc), a Newtonian solvent contribution of viscosity $\eta$, and an
isotropic pressure, $p$. Here $K_{\alpha\beta} =
\partial_\beta v_\alpha$ and $\tens{D}=\tfrac{1}{2}(\tens{K}
+\tens{K}^T)$. In considering only imposed displacements of the
(initial) form $\textbf{u}(\textbf{r},t)=u(y)\xhat=\gamma_0 y\xhat$,
we restrict all displacements and velocities to the direction $\xhat$
and all gradients to the direction $\yhat$.  Relevant associated
fields are then the displacement $u(y,t)$, strain
$\gamma(y,t)=\partial_y u(y,t)$, velocity $v(y,t)=\dot{u}(y,t)$ and
strain-rate $\gdot(y,t)=\partial_y v(y,t)$.  For the dynamics of the
elastoplastic stress $\boldsymbol{\sigma}$ we shall use two different
constitutive models. As suited to deformations of the form just
described, we track only the shear stress components $\sigma_{xy}$ and
$\Sigma_{xy}=\sigma_{xy}+\eta\gdot$ of the viscoelastic and total
stresses, further dropping the $xy$ subscript for clarity.

The first constitutive model to be used is the soft glassy rheology
(SGR) model~\cite{sollich1997rheology}. This considers an ensemble of
elements, each corresponding to a local mesoscopic region of material
(a few tens of emulsion droplets, say).  Under an imposed shear
deformation of rate $\gdot$, each element experiences a buildup of
local elastic shear strain $l$ according to $\dot{l}=\gdot$, with a
corresponding stress $Gl$, given a constant modulus $G$. This is
intermittently released by local plastic yielding events, each of
which is modelled as hopping of an element over a strain-modulated
energy barrier $E$, governed by a noise temperature $x$, with the
yielding intervals chosen stochastically with rate
$\tau_0^{-1}\exp[-(E-\tfrac{1}{2}kl^2)/x]$. Upon yielding, any element
resets its local stress to zero and selects its new energy barrier at
random from an exponential distribution $\exp(-E/x_{\rm g})$.  This
confers a broad spectrum of yielding times, $P(\tau)$. It also results
in a glass phase for $x<x_{\rm g}$, in which, in the absence of flow,
a material shows rheological ageing: the timescale for the relaxation
of the macroscopic stress $\langle Gl\rangle$ follows a step strain
increases linearly with the sample age $\tw$.  A steadily imposed
shear flow however interrupts ageing, and the steady state flow curve has a
yield stress. Full details of the SGR model in its original, spatially
uniform form are in Ref.~\cite{sollich1997rheology}.

The model's adaptation to account for non-uniform deformations is discussed
in~\cite{fielding2009shear,moorcroft2011age}. This involves
numerically taking $m=1...M$ SGR elements on each of $n=1...N$
streamlines at discretised flow-gradient positions $y=0...L_y$, with
periodic boundary conditions. The viscoelastic stress on streamline
$n$ is $\sigma_n = (G/M)\sum_m l_{nm}$.  Given an imposed average
shear rate $\gdotbar$ across the sample as a whole (which in our case
is zero at all times apart from $\tw$), the shear rate on each
streamline $n$ is then calculated, by enforcing force balance, to be
$\gdot_n=\gdotbar+(\bar\sigma-\sigma_n)/\eta$, where
$\bar\sigma=(1/N)\sum_n\sigma_n$.

\begin{figure}[!t]
  \includegraphics[width=8.5cm]{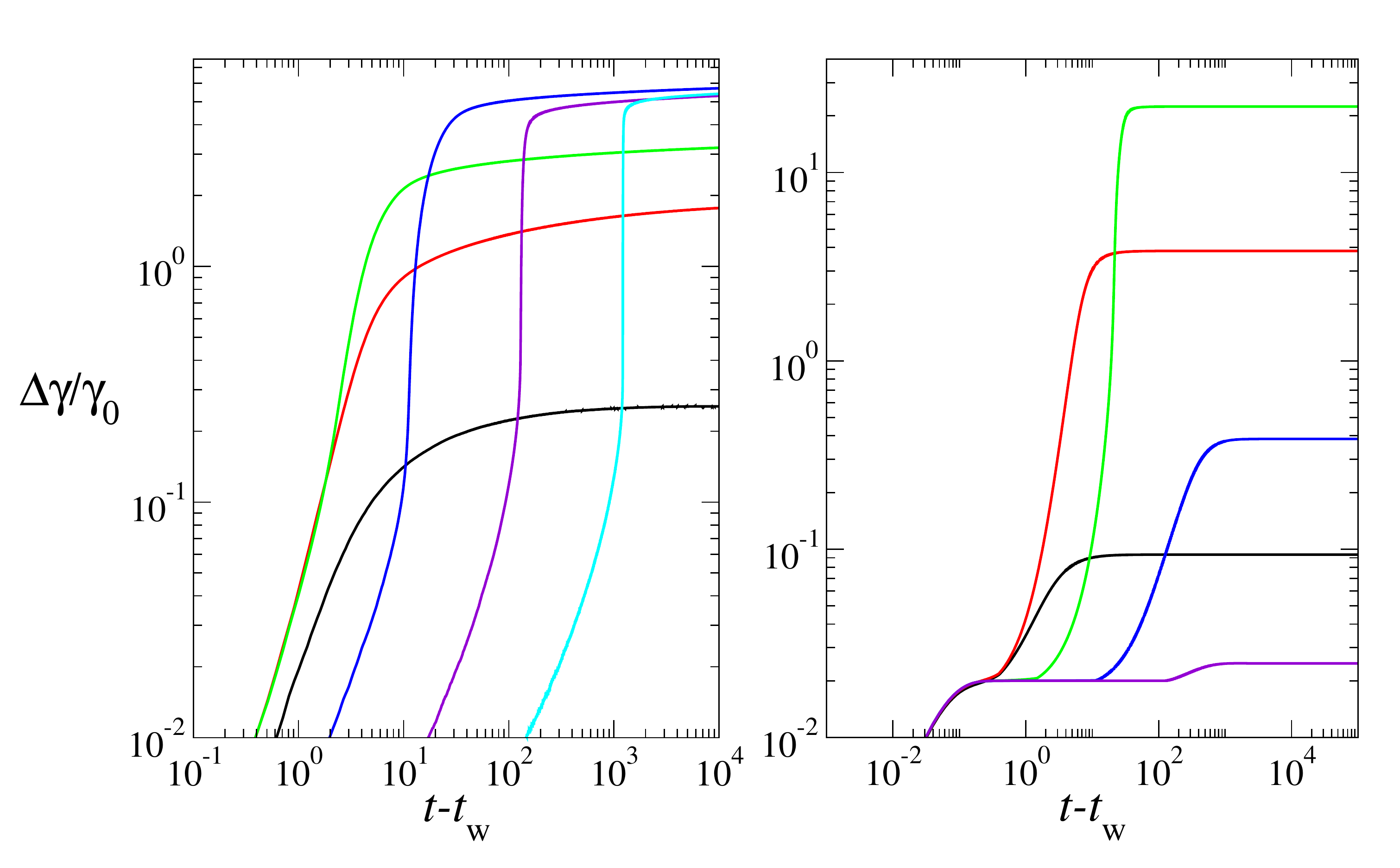}
  \caption{Degree of strain heterogeneity $\Delta\gamma$ across the
  sample, normalised by the imposed strain $\gamma_0$, as a function
  of the time interval $\Delta t=t-\tw$ since the imposition of a step
  shear strain, for a fixed strain amplitude $\gamma_0$ and waiting
  times $\tw$ as in Fig.~\ref{fig:fig1}. {\bf Left:} SGR model. {\bf Right:} fluidity model.}
\label{fig:fig2}
\end{figure}

Our second model is a highly simplified continuum fluidity
model~\cite{moorcroft2011age}, which supposes a Maxwell-type constitutive
equation for the viscoelastic stress
\be
\partial_t\sigma(y,t)=G\gdot-\sigma/\tau,
\label{eqn:sigma}
\ee
where $G$ is a constant modulus and $\tau$ is a structural relaxation
time (inverse fluidity) that has its own dynamics:
\be
\partial_t\tau=f(\tau,\sigma,\gdot)+l_o^2\partial^2_y \tau.
\label{eqn:tau}
\ee
We have considered three different model variants within this general
form. The first has $f=1-|\gdot|(\tau-\tau_0)(|\sigma|-\sigma_{\rm
th})\Theta(|\sigma|-\sigma_{\rm th})$ with $\sigma_{\rm th}=1$; the
second has $f=1-|\gdot|(\tau-\tau_0)$; the third has
$f=1-\tau/(\tau_0+1/|\gdot|)$. Each captures rheological aging, with
the timescale for stress relaxation following the imposition of a step
strain increasing linearly with the system age, $\tau =\tw$. A steady
flow cuts off ageing at the inverse strain rate, and the steady state
flow curve displays a yield stress.  The results that we present below
are all obtained within the first functional form for $f$, but we have
checked that the same scenario qualitatively holds within all three
variants.  The parameter $l_o$ in Eqn.~\ref{eqn:tau} is a mesoscopic
length describing the tendency for the relaxation time of a
mesoscopic region to equalise with those of its neighbours.

\begin{figure}[!t]
  \includegraphics[width=8.5cm]{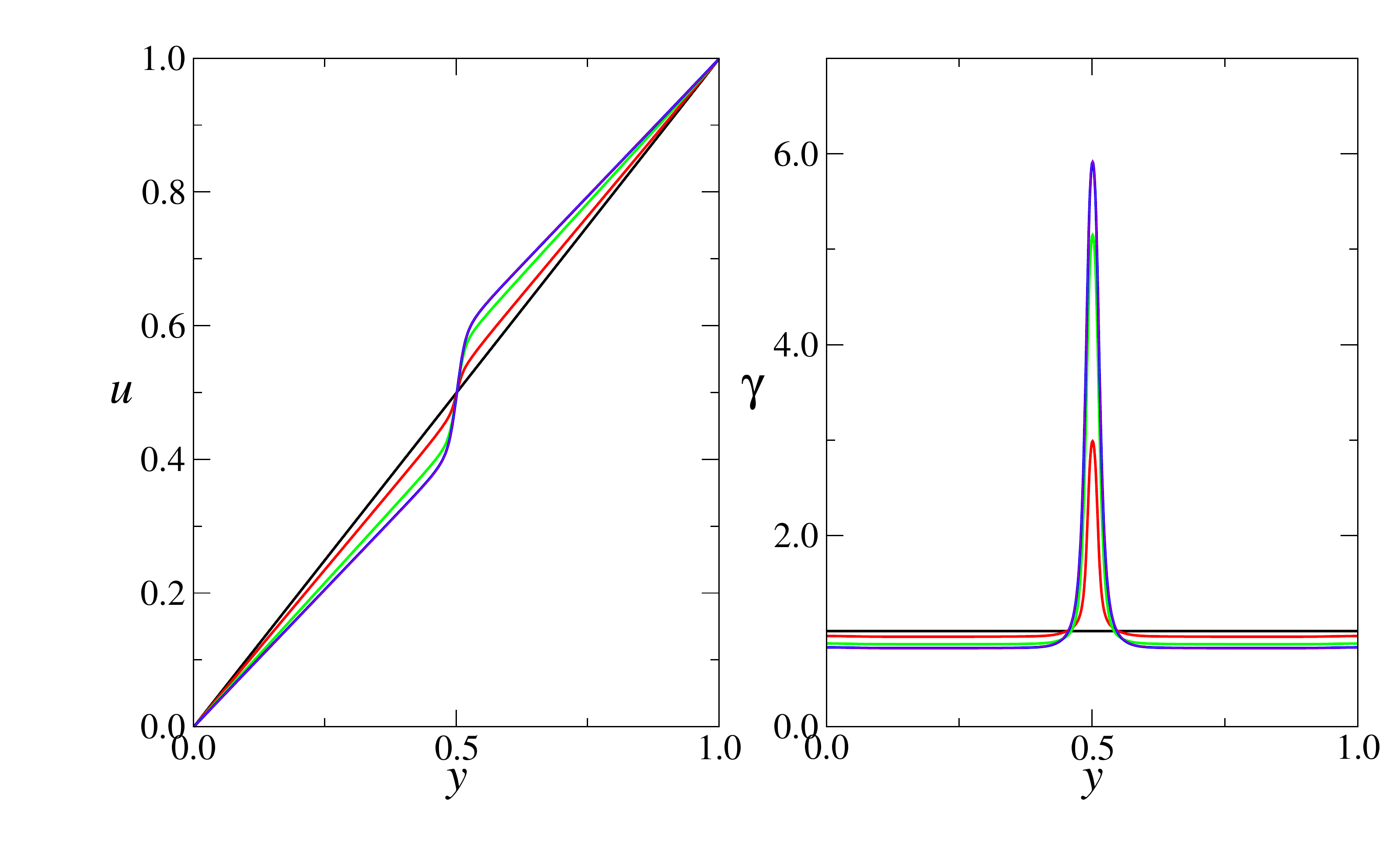} \caption{{\bf
  Left:} displacement as a function of position across the sample at
  time intervals $\Delta t=0.0$ (black), $40.0$ (red), $45.0$ (green),
  $60.0$ (blue) and $80.0$ (violet) following the imposition of a
  strain of amplitude $\gamma_0=7.0$ at a waiting time $\tw=10^{10}$
  in the fluidity model. {\bf Right:} corresponding strain
  field. Profiles for $\Delta t=60.0$ and $80.0$ are indistinguishable.} 
  \label{fig:fig3}
\end{figure}

Within each of these constitutive models, we shall consider a slab of
material sandwiched between infinite flat parallel plates at $y=0,
L_y$. We assume it to be freshly prepared at time $t=0$ in a fully
rejuvenated initial state with zero stress across the whole sample,
$\sigma(y,t=0)=0$.  The sample is then left to age undisturbed until a
time $\tw$, when it is suddenly subject to an (initially) uniform
shear deformation $\textbf{u}(\textbf{r},t)=\gamma_0 y\xhat$, by
displacing the top plate relative to the bottom one a distance
$\gamma_0 L_y$ in the positive $\xhat$ direction, generating a stress
$\sigma(t=\tw^+)=G\gamma_0$.  (In reality, inertia requires a non-zero
time to accomplish this; as noted above, the scenario we present holds
for any short deformation interval.) No further (global) strain is
imposed thereafter, with zero average shear rate across the sample
$\gdotbar
\equiv
\int_0^{L_y}\gdot(y,t)dy =0$.

As a function of the subsequent time $\Delta t=t-\tw$, we track the
decay back to zero of the (total) shear stress $\Sigma(\Delta t)$. We
also track the displacement field $u(y,\Delta t)$ and associated
strain field $\gamma(y,\Delta t)=\partial_y u(y,\Delta t)$ within the
sample. Note that the (initially uniform) strain field
$\gamma(y,\Delta t)$ can become heterogeneous across the sample
($y$-dependent) as a result of the instability that we report, whereas
$\Sigma$ must remain uniform by force balance. To seed the
instability, we add a small initial heterogeneity across the
sample. We have checked that the scenario we present is robust to the
nature and size of this.

We rescale strain, stress, time and length so that $x_{\rm g}=G=\tau_0=L_y=1$.
The solvent viscosity $\eta$ is expected to be much less than the
viscosity scale of the viscoelastic component, $G\tau_0=1$, but is
otherwise unimportant to the physics we describe.  For simplicity we
set $\eta=0.05$, but have checked for robustness in variations in this.

\begin{figure}[!t]
  \includegraphics[width=0.85\columnwidth]{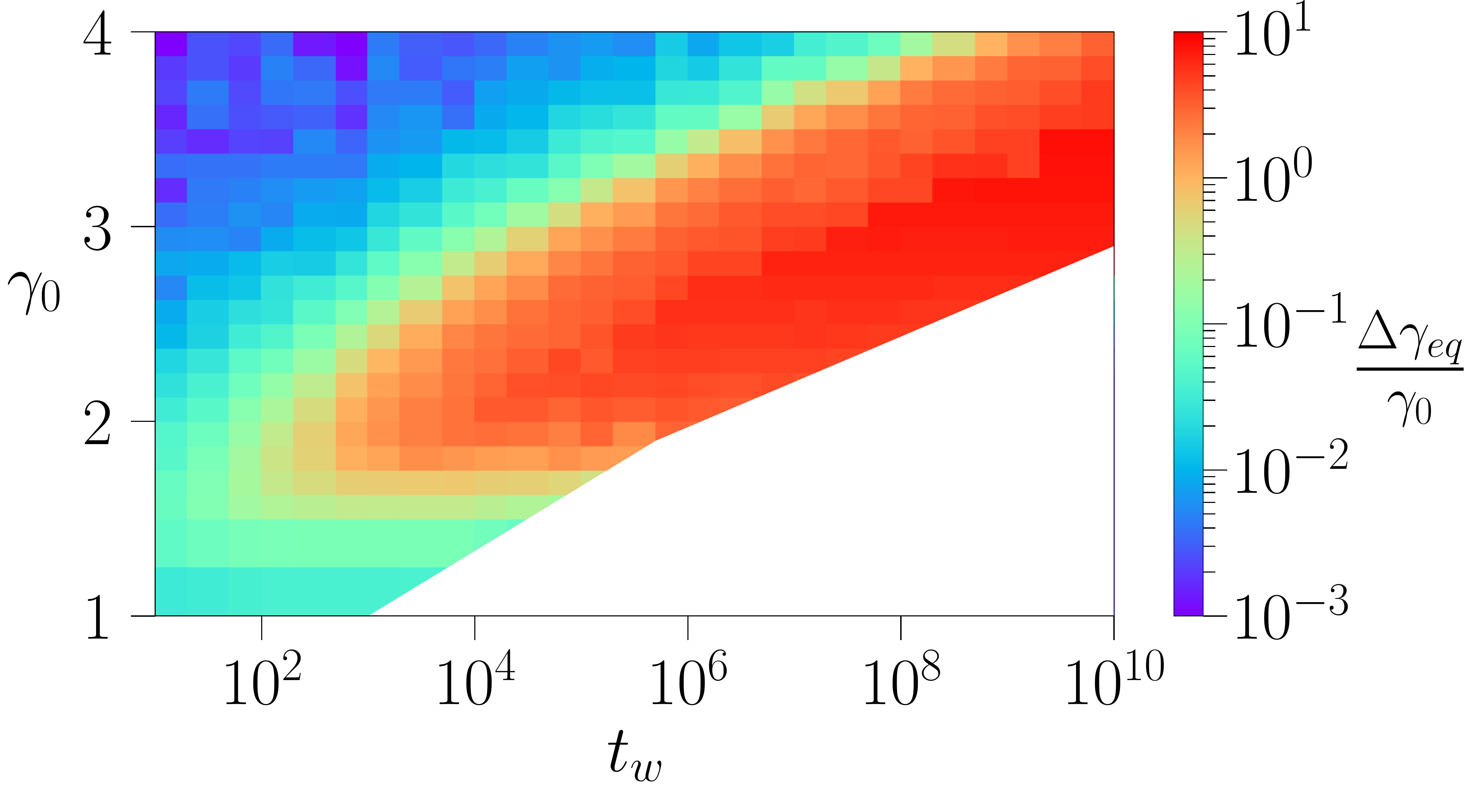}
  \includegraphics[width=0.85\columnwidth]{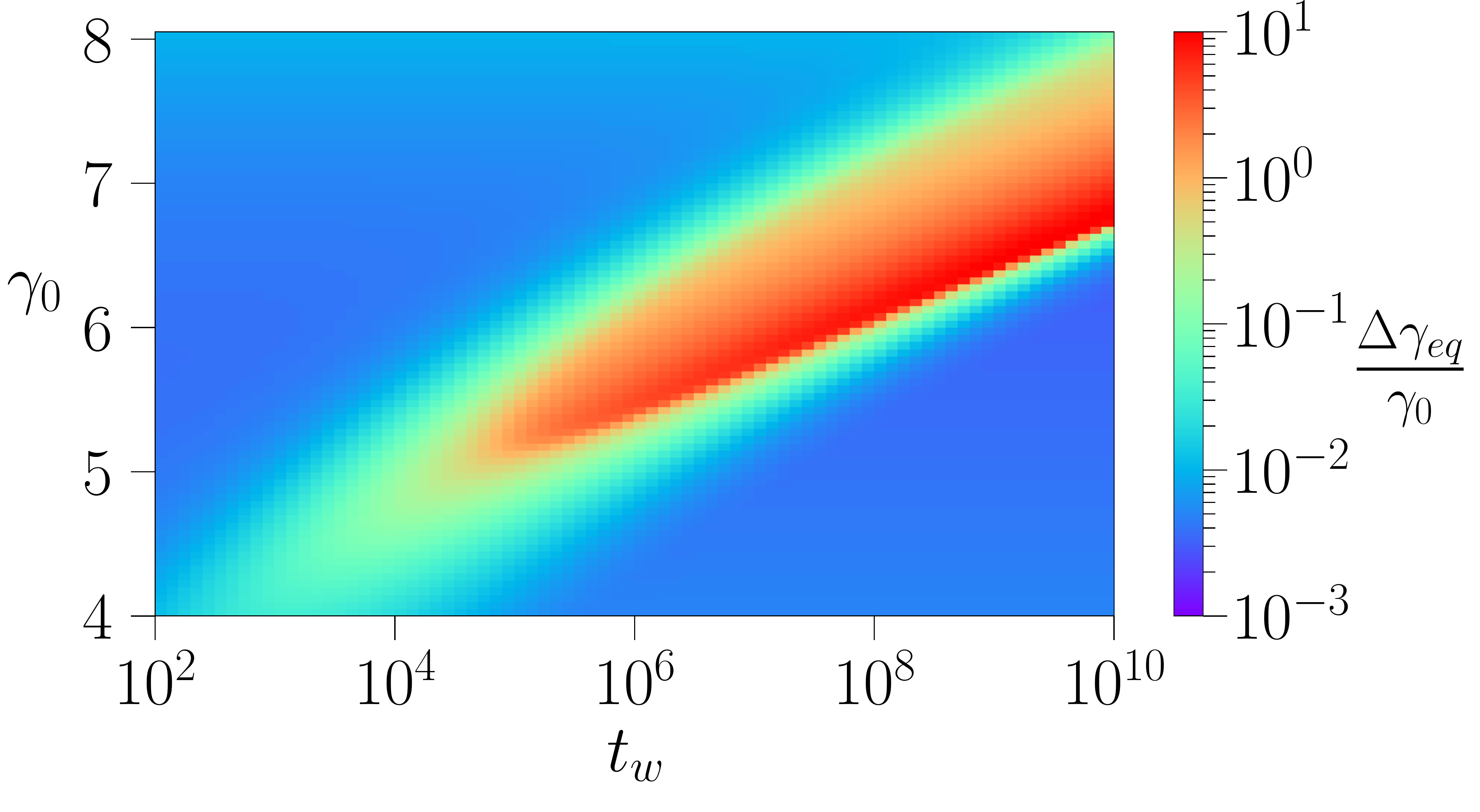}
  \caption{Colourmap showing the normalised degree of strain
  heterogeneity $\Delta \gamma/\gamma_0$ attained at long times after
  the imposition of a step strain as a function of the amplitude
  $\gamma_0$ of the imposed strain and the sample age $\tw$ at the
  time the strain was applied. {\bf Top:} in the SGR model at $x=0.3$.
  In the region shown in white, run-times are too long to obtain
  results across a full phase diagram, although slices (of the time at
  which the instability occurs) are shown to $\tw=10^{10}$ for several
  $\gamma_0$ in Fig.~\ref{fig:fig5}. {\bf Bottom:} in the fluidity
  model.}  \label{fig:fig4}
\end{figure}

Fig.~\ref{fig:fig1} shows the stress decay in the SGR model (left),
and fluidity model (right), following the imposition of a step strain
of a fixed amplitude, $\gamma_0$, for several different sample ages
$\tw$. In each case, the timescale of stress decay increases with the
sample age at the time the strain is imposed. The dashed lines show
the results of calculations in which the strain field $\gamma(y)$ is
artificially constrained to remain homogeneous during the stress
decay, independent of $y$. The solid lines show calculations in which
it is allowed by become heterogenous across the sample. The departure
of the latter from the former marks the onset of an instability in
which the strain field  becomes heterogeneous, accompanied by a
more precipitous drop in the stress signal than is predicted by the
artificially constrained homogeneous calculation.

\begin{figure}[!t]
  \includegraphics[width=8.5cm]{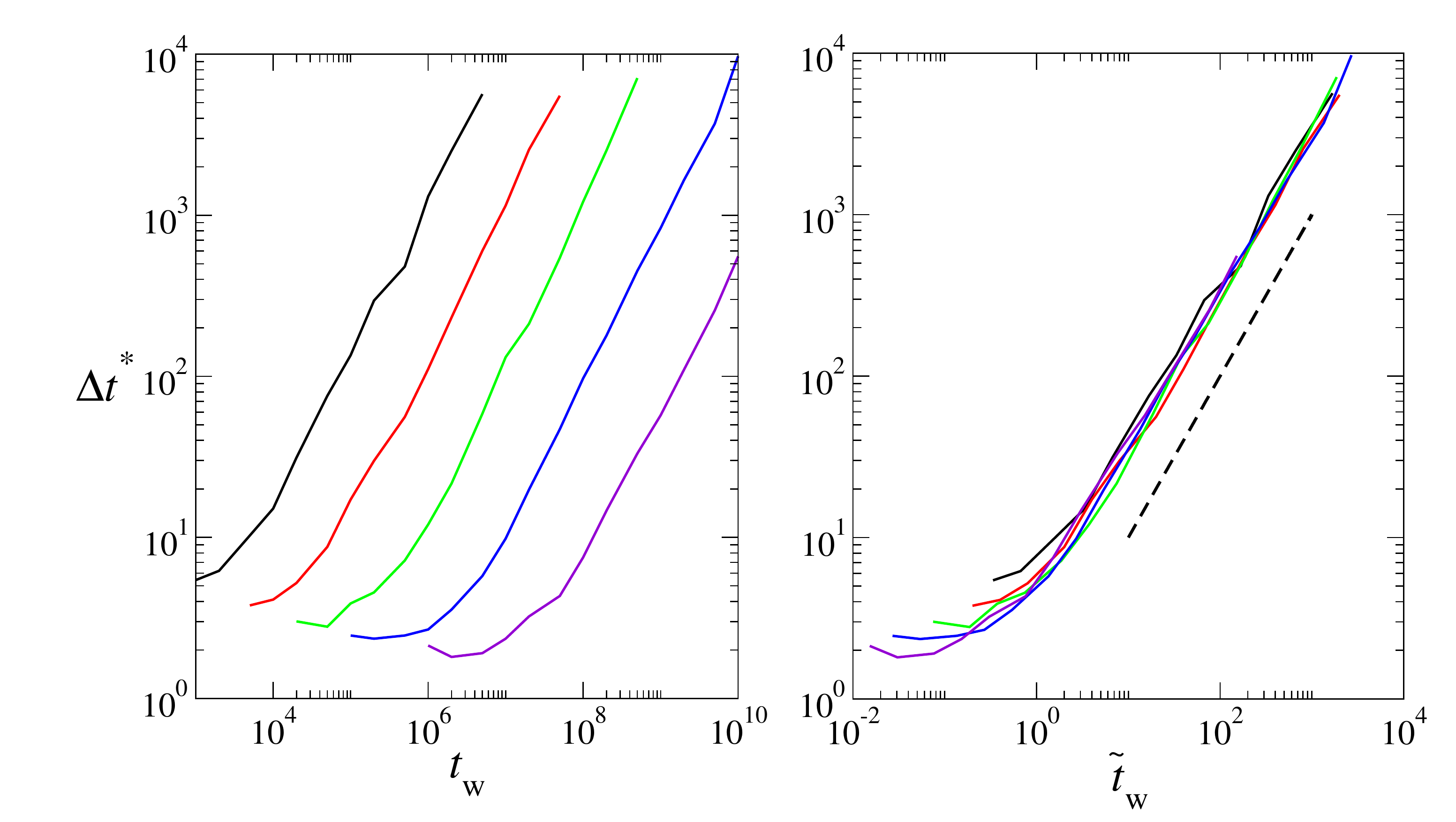}
  \caption{{\bf Left:} time interval $\Delta t^*$ following the imposition of a step strain  at which the strain localisation instability arises, as calculated within the SGR model. Results are plotted as a function of the sample age at the time the strain is imposed, for different values of the strain amplitude: $\gamma_0= 2.0$, $2.25$, $2.5$, $2.75$, $3.0$ in curves  left to right. {\bf Right:} same data replotted as a function of the scaled time $\tilde{t}_{\rm w}=\tw\exp(-\alpha\gamma_0^2/2x)$ with $\alpha\approx 1.2$. Dashed line shows $\Delta t^*=\tilde{t}_{\rm w}$ as a guide to the eye.}
  \label{fig:fig5}
\end{figure}

To characterise the degree of growing heterogeneity, we define the
quantity $\Delta\gamma(\Delta t)$ as the difference at any time
$\Delta t$ post-strain between the maximum of the strain
$\gamma(y,\Delta t)$ across the flow gradient direction
$y$, and the correspondingly defined minimum strain. The
time-evolution of this quantity is shown in Fig.~\ref{fig:fig2}, with
parameter values and line colours corresponding to those of the stress
decay in Fig.~\ref{fig:fig1}. The divergence of the stress decay
curves between the (enforced to be) homogeneous and (allowed to be)
heterogeneous runs in Fig.~\ref{fig:fig1} indeed arises contemporarily
with the formation of a heterogeneous strain field, as characterised
by the growth of $\Delta
\gamma(t)$ in Fig.~\ref{fig:fig2}. At long times the strain heterogeneity settles to a constant in
the fluidity model. (In the SGR model, it continues a very slow
logarithmic growth due to the small noise present our stochastic
simulations, which however decreases with increasing number of
elements $M$.)

For one particular value of imposed strain amplitude $\gamma_0$ and
waiting time $\tw$, we show in Fig.~\ref{fig:fig3} the displacement
field $u(y,\Delta t)$ and the strain field $\gamma(y,\Delta t)$, at
several time intervals $\Delta t$ following the imposition of the
strain, as calculated within the fluidity model.  As can be seen, the
initially linear displacement field $u(y,\Delta t=0)=\gamma_0 y$ gives
way to a non-linear one, associated with a pronounced heterogeneity in
the strain field $\gamma(y,\Delta t)$, consistent with the observed
temporal growth in $\Delta
\gamma(\Delta t)$. The same quantities computed in the SGR model display the same behaviour (not shown).

We discussed in Fig.~\ref{fig:fig2} the growth in strain heterogeneity
$\Delta\gamma(\Delta t)$ as a function of the time $\Delta t$
post-strain, for one fixed value of the imposed strain amplitude
$\gamma_0$ and several values of the sample age $\tw$.  In
Fig.~\ref{fig:fig4}, we summarise in a `phase diagram' colourmap the
limiting degree of strain heterogeneity $\Delta\gamma$ at long times
in the full plane of $\gamma_0,\tw$. (In practice, we take this value
at the final time of the run for the fluidity model, and just after
the precipitous rise in this quantity in the SGR model, to cutoff the
slow logarithmic growth still present at long times in the SGR model.)
Each coordinate pair $(\gamma_0,\tw)$ in this plane corresponds to a
single step strain experiment, with the colourscale showing the final
degree of strain heterogeneity, normalised by the amplitude of the
initially imposed strain: $\Delta
\gamma/\gamma_0$. A significant degree of
strain heterogeneity is observed across large regions of this parameter
space, in both the SGR model (top) and fluidity model (bottom). In the
region shown as white for the SGR model, run times are too long to
obtain results.

We discuss finally the time delay $\Delta t^*$ after the imposition of
the strain at which the instability arises. We plot this in
Fig.~\ref{fig:fig5} (left) as a function of the sample age $\tw$ for
several values of the imposed strain amplitude $\gamma_0$, in the SGR
model. The same data are replotted in the right panel as a function of
the scaled time $\tilde{t}_{\rm w}=\tw\exp(-\alpha\gamma_0^2/2x)$,
showing good data collapse onto a line $\Delta t^*\propto \tilde{t}_{\rm w}$. The
time at which the instability sets in can therefore be delayed long
into the process of stress relaxation, with the delay time becoming
arbitrarily long for initially old samples $\tw\to\infty$.  It is
worth emphasising this remarkable finding: that a catastrophic
instability can arise within a material at indefinitely long times
after any external deformation was last applied.

\iffalse
One obvious difference in the predictions of the two models, for which
we have no explanation to date, is that the regime of significant
instability is cutoff at the lower right of the $\gamma_0,\tw$ plane
by a diagonal phase boundary in the fluidity model (the same holds in
the other two fluidity model variants, not shown), but by a horizontal
boundary in the SGR model: at a fixed $\gamma_0$, the instability is
lost in the fluidity model by taking a large enough initial sample age
$\tw$, but persists even as $\tw\to\infty$ in the SGR model. Because
the SGR model is a more sophisticated model that has been shown to
capture the rheology of yield stress fluids in numerous experimental
protocols, we suggest it to provide the better description.
\fi

A banding instability after a rapid shear strain has been observed
previously in polymer melts~\cite{boukany2009step,fang2011shear},
although after a short delay time of just a few seconds (consistent
with the absence of long-term memory in those ergodic fluids), and
having its origin in a non-monotonic relationship between stress and
strain during the initial rapid straining
process~\cite{agimelen2013apparent,moorcroft2014shear}. No such
non-monotonicity exists in any model explored here for an infinite
rate of strain imposition, suggesting a fundamentally different
instability mechanism in these soft glassy materials. Indeed, we
suggest the mechanism to be as follows. Imagine an initially near
uniform sample, but with a streamline (or region) in which the strain
increases slightly relative to the rest of the sample. This slightly
fluidises the material on that streamline, causing the elastoplastic
stress $\sigma$ to relax slightly faster. To maintain a uniform total
stress $\Sigma$, that streamline must strain forward slightly
further. This represents a positive feedback loop, leading to a
runaway instability.

To summarise, we have uncovered a strain localisation instability that
arises at long times after the imposition of a step strain in soft
glassy materials, accompanied by a precipitous drop in the shear
stress.  We have explored the phenomenology of this instability via
detailed numerical simulations of a mesoscopic soft glassy rheology
model, and three different variants of a highly simplified continuum
fluidity model. In finding the basic features to be the same across
these different constitutive models, we suggest that the instability
reported here may be generic across amorphous, glassy materials. We
hope that these predictions will stimulate experimental studies aimed
at observing this instability. A particularly remarkable feature is
that the instability can arise at extremely long times after the
initial strain imposition, i.e., long after the material last suffered
any mechanical deformation, due to the long-term memory inherent to
glassy materials. This could have far reaching consequences for
material processing and performance, and for delayed geophysical
phenomena such as seismic aftershocks.

\bibliographystyle{apsrev2}
\bibliography{refs}

\end{document}